\documentclass[11pt]{article}
\usepackage[margin=1in]{geometry}
\usepackage{amsmath,amssymb,amsthm,booktabs,array}
\usepackage[authoryear,round]{natbib}
\usepackage[colorlinks=true,allcolors=blue]{hyperref}
\usepackage{graphicx,microtype,placeins}
\graphicspath{{figures/}}
\hypersetup{pdftitle={Certified computation of Buehler confidence limits for system reliability},pdfauthor={Zhaohui Li}}
\newtheorem{theorem}{Theorem}
\newtheorem{proposition}{Proposition}
\newtheorem{corollary}{Corollary}
\newcommand{\E}{\mathbb E}
\newcommand{\Prb}{\mathbb P}
\newcommand{\rvec}{\mathbf r}
\newcommand{\one}{\mathbf 1}
\newcommand{\sig}{\operatorname{sigmoid}}
\title{Certified computation of Buehler confidence limits\\for system reliability}
\author{Zhaohui Li\\[0.4em]
\small State Key Laboratory of Mathematical Sciences (SKLMS),\\
\small Academy of Mathematics and Systems Science,\\
\small Chinese Academy of Sciences, Beijing 100190, China\\[0.3em]
\small\href{mailto:lizh@amss.ac.cn}{\texttt{lizh@amss.ac.cn}}}
\date{31 October 2026}
\begin{document}
\maketitle
\begin{abstract}
Component tests can support a lower confidence limit for a system that has not been tested as a whole, but numerical optimization can compromise its coverage.
A feasible point in the minimization defining a Buehler limit gives an upper bound on that limit; reporting it can therefore overstate reliability.
We develop a computation that maintains lower and upper bounds on the ideal endpoint for a prescribed ordering.
Bounds over entire parameter regions justify the reported lower endpoint at every completed checkpoint, including under adaptive stopping and selection among runs.
Separate refinement conditions establish convergence, and an explicit error allocation accounts for simulated tail probabilities and confidence-region localization.
The theory retains nuisance and dependence parameters in the observation model.
Monotone corner bounds give the implemented specialization for scalar and two-component binomial systems; Bernstein and probability-distance bounds provide further theoretical constructions.
Four benchmark problems reach a certified width of $10^{-5}$ under each of four search strategies.
Additional stochastic candidates increase cost without reducing subdivisions, whereas complete sample-space enumeration verifies the implemented coverage at three parameter vectors.
For the all-success example, the certified lower limit is $0.68128$, compared with $0.41738$ from Bonferroni-adjusted component limits.

\end{abstract}
\noindent\textbf{Keywords:} system reliability; Buehler confidence limits; numerical certification; global optimization; nuisance parameters.

\section{Introduction}
\label{sec:introduction}
Component test data often provide the only direct evidence about the reliability of a system whose complete testing would be costly.
The inferential problem is to combine those observations with the system structure and report a lower limit that accounts for sampling uncertainty.
For example, with six and eight successful tests of two series components, the $90\%$ Bonferroni component-product limit is $0.41738$, whereas the ordered construction studied here has ideal endpoint $0.68129$.
Computing the latter accurately is useful only if the returned number preserves its coverage guarantee.

Buehler's ordered confidence construction expresses this assessment through a tail probability and a constrained optimization \citep{buehler1957,lloyd2003}.
The construction is applicable to a general model with nuisance parameters.
In a reliability problem, the target can be a function of component reliabilities, lifetime parameters and parameters describing dependence.
The designated statistic orders the possible datasets, and the lower limit is the smallest reliability among parameter values satisfying the tail constraint.
Its finite-sample coverage is a statistical property of this inversion.

The numerical calculation introduces another question.
A feasible solution of the minimization gives an upper bound on the ideal lower confidence limit.
Reporting that objective value may therefore increase the limit beyond the value justified by the ordered construction.
The computation needs a global lower bound on the optimization problem to determine what can be reported at the current budget.

In this paper, we develop a computation that maintains this lower bound together with an upper bound on the ideal endpoint.
The parameter space is covered by cells.
Objective bounds and tail-probability bounds allow the algorithm to exclude cells that cannot contain an improving feasible point.
The retained cells determine the reportable lower endpoint.
The interval width bounds the remaining numerical error in the endpoint.
Candidate searches can improve the upper bound, but the difficult step is to raise the global lower bound while retaining every region that could still contain a better feasible point.

The method is formulated in terms of a full observation law, a reliability functional and a prescribed ordering statistic.
This separates the statistical model from the particular representation used to compute bounds.
Monotonicity gives inexpensive bounds for coherent systems.
Bernstein subdivision supplies bounds for nonmonotone binomial events and arbitrary binary structures.
Probability-distance bounds provide another route for continuous lifetime models, with nuisance parameters retained in the search.

We develop four connected results.
First, valid cell enclosures yield a confidence limit at every finite completed computation.
Second, this property is preserved under adaptive stopping and selection among candidates or completed runs for the same prescribed ordering.
Third, enclosure consistency and refinement conditions establish convergence of the numerical interval.
Fourth, an explicit error allocation combines original-data uncertainty, simulated probability bounds and data-dependent localization of nuisance parameters.
The numerical study evaluates the exact monotone-binomial implementation on four scalar and two-component problems; the nonmonotone, continuous-model and simulation-based constructions are theoretical extensions whose computational performance is not evaluated here.

The paper is organized as follows.
Section~\ref{sec:literature} reviews reliability inference, ordered limits and the computational methods most closely related to this work.
Section~\ref{sec:framework} defines the statistical target.
Section~\ref{sec:algorithm} develops the general certificate and its guarantees.
Section~\ref{sec:enclosures} gives model-specific enclosures, and Section~\ref{sec:candidates} describes optional candidate generation.
Section~\ref{sec:numerical} reports the numerical study.
Section~\ref{sec:conclusion} discusses the resulting method and further computational development.

\section{Related work}
\label{sec:literature}
\subsection{Reliability assessment from component data}
Likelihood-based reliability synthesis combines component information with a constraint on the system reliability.
\citet{madansky1965} treated binomial series, parallel and series--parallel configurations, while approximations for exponential series systems and mixed binomial--exponential data were developed by \citet{elmawaziny1967} and \citet{manngrubbs1974}.
The assessments of \citet{oneill1972,winterbottom1984} place these procedures within the broader problem of interval estimation from component tests.
Their common computational task is to combine a system structure with an observation model; their approximate calibration leaves a separate question about finite-sample coverage.

Exact procedures exploit distributional identities supplied by the sampling design.
Gamma-parameter inference provides one route \citep{lentner1963}, while transformations and conditioning give exact limits for exponential series systems \citep{sarkar1971,lieberman1971}.
Negative-binomial sampling supports exact limits for functions of Bernoulli parameters \citep{hwang1973}; more general system distributions and quantiles can be treated by asymptotic procedures, including the implicit-delta approach of \citet{hong2014}.
These distinctions matter computationally: a pivot can simplify the tail calculation, whereas a general observation law may require integration or simulation with nuisance parameters retained.

\subsection{Ordering, nuisance parameters and Buehler computation}
Buehler's original construction concerns a product of two binomial parameters \citep{buehler1957}, but its designated statistic is a substantive inferential choice.
Calibrating an approximate limit while preserving its ordering \citep{kabaila1997} and choosing a statistic that gives tight limits \citep{kabaila1998} address different parts of this choice.
Approximate-limit orderings can reduce large-sample conservatism, and the designated statistic also affects system-failure bounds \citep{kabaila2001,kabaila2002}.
Subsequent studies of efficiency, refinement and nesting show why an ordering cannot be treated as a numerical tuning parameter \citep{kabaila2003,kabaila2006,kabaila2004}; monotonicity of likelihood-support bounds raises a related issue \citep{lloyd2005}.
We therefore fix the ordering before computation and measure numerical error against its particular endpoint.

The general formulation and ordering-specific optimality in \citet{lloyd2003} already include nuisance-parameter optimization.
Applications to relative risks and odds ratios \citep{wangshan2015}, average heterogeneous Bernoulli success probabilities \citep{mattner2015}, and sequential-trial decisions \citep{lloyd2022} demonstrate the breadth of ordered inference beyond reliability products.
Our statistical target uses this established formulation; the question is how to bound its numerical approximation.

Existing computational approaches attack complementary costs: EM handles nuisance parameters and incomplete data \citep{fang2005}, importance sampling estimates parameter-indexed tail surfaces \citep{lloydli2014}, and constrained optimization supports complex-system reliability assessment from component data \citep{du2020}.
However, an approximate tail surface or a feasible optimizer alone does not give the global lower bound needed to report a conservative endpoint.
The comparison relevant here is the cost of enclosing a prescribed endpoint to a specified width, which also requires accounting for tail-evaluation error and unexplored parameter regions.
The numerical comparison below isolates the cost of optional candidate search within a common certifier; comparison with these earlier computational methods would require matching both the statistical target and the error guarantee.

\subsection{Global bounds and optional search}
Whole-region bounds are standard tools in global optimization: convex underestimators handle factorable nonconvex programs \citep{mccormick1976}, and interval methods enclose multivariate objectives \citep{hansen1980}.
Complete search combines such bounds with propagation, local searches and subdivision \citep{neumaier2004,csendesratz1997}.
Here a good feasible point lowers the incumbent, whereas excluding a region requires a bound valid at every parameter in that region.
Only the latter operation can justify raising the reported lower confidence limit.

For binomial observations, a fixed count-space event already has a tensor-product Bernstein representation.
Polynomial arithmetic and coefficient range bounds therefore supply a direct route to probability enclosures, including for nonmonotone events \citep{faroukirajan1987,faroukirajan1988,garloff1993}.
Variation and divergence inequalities offer a different route for continuous models by controlling probability changes across a parameter cell \citep{csiszar1967,sasonverdu2016,csiszarshields2004}.
The practical tradeoff is enclosure sharpness against evaluation and storage cost.

Stochastic approximation \citep{robbins1951,nemirovski2009} and likelihood-ratio gradients \citep{glynn1990} can propose parameters near the tail constraint.
Nonconvex stationary-point guarantees \citep{ghadimi2013} and methods for expectation constraints \citep{lanzhou2020} support this search, as do sample-average and conservative approximations to probability constraints \citep{luedtkeahmed2008,nemirovskishapiro2006}.
Their role in our algorithm is optional: although a candidate can reduce the upper bound, it is accepted only after a valid feasibility check, and its search cost must be included in the comparison.

\subsection{Simulation error and confidence-set inversion}
Sequential Monte Carlo $p$-values and resampling-risk control make simulation error part of the inferential calculation \citep{besagclifford1991,gandy2009}.
Time-uniform concentration, confidence sequences and betting methods permit adaptive probability estimation and testing \citep{howard2020,howard2021,waudbysmithramdas2024,fischerramdas2025}.
For parameter-space search, such statements must also cover every simulation call used to accept a candidate or exclude a cell.

\citet{glazerstark2026} invert valid Monte Carlo tests using shared auxiliary randomness, develop conservative numerical inversion, and discuss nuisance parameters and projections onto nonlinear functionals.
Their target is the randomized test being inverted; in our exact-enclosure branch, the target is the population-tail Buehler endpoint, and the interval width bounds numerical error relative to that endpoint.
Our simulated-enclosure branch instead reserves an explicit probability of computational failure.
Confidence-set localization follows the error-allocation principle of \citet{bergerboos1994}, who maximize a $p$-value over a nuisance-parameter confidence set and add its noncoverage allowance.

The method below maintains a lower bound on the constrained minimum together with an upper bound from certified feasible points, so their gap bounds how far the reported limit can fall below the ideal endpoint.
Its finite-budget argument uses whole-cell inequalities and is unchanged when corner, Bernstein or probability-distance bounds supply them.
This separates the coverage argument from the choice of search routine, while making the cost and sharpness of the enclosures the central computational questions.

\section{Statistical formulation}
\label{sec:framework}
\subsection{System reliability and the observation model}
Suppose that the observed test data $X$ have distribution $P_\theta$, where $\theta\in\Theta$ is the full parameter vector.
The data may contain component success counts, observed lifetimes, censoring indicators or joint system states.
We write the reliability of interest as
\[
R=R(\theta),\qquad 0\le R(\theta)\le1.
\]
The parameter $\theta$ includes all quantities needed to specify the observation law and system reliability, including distributional shape and dependence parameters.
The statistical formulation therefore accommodates different component models and system structures through $P_\theta$ and $R(\theta)$.

For independent binary component states, let $Z_i$ indicate whether component $i$ functions and put $\Prb_\theta(Z_i=1)=r_i$.
A binary structure function $\varphi:\{0,1\}^d\to\{0,1\}$ gives
\begin{equation}
R(\theta)=f(\rvec)
=\sum_{z\in\{0,1\}^d}\varphi(z)
 \prod_{i=1}^d r_i^{z_i}(1-r_i)^{1-z_i}.
\label{eq:system}
\end{equation}
The function $\varphi$ may describe a series, parallel, voting, bridge or other binary structure.
For dependent states with joint probabilities $\pi_\theta(z)$, the corresponding expression is
\begin{equation}
R(\theta)=\sum_{z\in\{0,1\}^d}\varphi(z)\pi_\theta(z).
\label{eq:dependent}
\end{equation}
Dependence parameters remain in $\theta$ when the component marginal reliabilities do not determine these joint probabilities.
For multistate systems or continuous performance measures, $R(\theta)$ can instead be the probability that a specified system performance criterion is met.

Let $T:\mathcal X\to\mathbb R$ be a prescribed statistic, with larger values representing stronger evidence for higher reliability.
A natural choice is the plug-in system estimate $T(X)=R(\widehat\theta(X))$.
The statistic can include a prescribed fitting or learning procedure.
The same procedure is applied to the observed data and every replicated dataset used to evaluate its distribution.
Thus nuisance parameters estimated in $T$ are refitted in each replicate.

For observed data $x$, define
\begin{equation}
G_x(\theta)=\Prb_\theta\{T(X')\ge T(x)\},
\label{eq:tail}
\end{equation}
where $X'$ follows the full observation model, including the censoring or sampling design.
The inclusive tail treats ties in discrete and mixed distributions.
At significance level $\alpha\in(0,1)$, the ideal ordered lower limit is
\begin{equation}
L_\alpha(x)=\inf_{\theta\in\Theta:\,G_x(\theta)>\alpha}R(\theta),
\label{eq:ideal}
\end{equation}
with the value one when the feasible set is empty.
All statistics, limits and computational outputs below are assumed measurable.

\subsection{The ordered confidence construction}
The following result states the classical ordering argument in the notation needed for the computational method \citep{buehler1957,lloyd2003}.
\begin{theorem}[Ordered lower confidence limit]
\label{thm:ideal}
For every $\theta_0\in\Theta$,
\[
\Prb_{\theta_0}\{L_\alpha(X)\le R(\theta_0)\}\ge1-\alpha.
\]
Moreover, $L_\alpha(x)$ is nondecreasing in $T(x)$ and is the largest nonrandomized lower confidence limit of the form $h(T(x))$ among nondecreasing functions $h:\operatorname{range}(T)\to[0,1]$ satisfying this coverage inequality for every $\theta_0$.
\end{theorem}
\begin{proof}
For any real random variable $Y$, its inclusive survival function $S(y)=\Prb(Y'\ge y)$ satisfies
\[
\Prb\{S(Y)\le a\}\le a,\qquad 0\le a\le1.
\]
Indeed, the set $\{y:S(y)\le a\}$ is an upper interval.
If its endpoint is included, its probability is the survival probability at that endpoint.
If the endpoint is excluded, take a decreasing sequence of included points and use continuity of probability.
The same argument applies to discrete, continuous and mixed laws.

When $G_X(\theta_0)>\alpha$, the true parameter belongs to the set in \eqref{eq:ideal}, so $L_\alpha(X)\le R(\theta_0)$.
Consequently, noncoverage is contained in $\{G_X(\theta_0)\le\alpha\}$, whose probability is at most $\alpha$.
Increasing $T(x)$ decreases $G_x(\theta)$ for every $\theta$, and hence increases the infimum in \eqref{eq:ideal}.

For maximality, suppose that $G_x(\theta)>\alpha$ and $h(T(x))>R(\theta)$.
Monotonicity of $h$ implies noncoverage throughout the event $\{T(X)\ge T(x)\}$, which has probability greater than $\alpha$.
This contradicts the assumed coverage of $h$.
Therefore $h(T(x))\le R(\theta)$ for every strict-feasible $\theta$.
Taking the infimum proves the assertion; for an empty feasible set, $h\le1$ gives the same result.
\end{proof}

The theorem defines the ordered target for the full observation model, including dependent components and nuisance parameters.
The following sections compute this target with a finite-budget guarantee.

\subsection{The direction of numerical approximation}
A strict-feasible parameter vector $\widetilde\theta$ gives
\begin{equation}
G_x(\widetilde\theta)>\alpha
\quad\Longrightarrow\quad
L_\alpha(x)\le R(\widetilde\theta).
\label{eq:primaldirection}
\end{equation}
Thus a candidate solution supplies an upper bound on the optimal value.
For confidence reporting, we also need a lower bound $\ell(x)\le L_\alpha(x)$.
Together these quantities give
\[
\ell(x)\le L_\alpha(x)\le u(x).
\]
The lower endpoint is the reported confidence limit, while $u-\ell$ bounds its numerical distance from the ideal limit.

Consider independent binomial observations with $n=(6,8)$, $x=(6,8)$ and $\alpha=0.1$ for a two-component series system.
Here
\[
G_x(r_1,r_2)=r_1^6r_2^8\le(r_1r_2)^6,
\qquad
L_\alpha(x)=0.1^{1/6}=0.681292069057\ldots.
\]
The vector $(0.85,0.85)$ is strict-feasible because $0.85^{14}\approx0.10276967$, and its objective value is $0.7225$.
If a procedure reports this value for the all-success observation, it exceeds the true reliability $0.72$ at $\rvec_0=(0.8,0.9)$.
That observation alone has probability $0.8^6 0.9^8=0.11284439629824$.
The coverage of such a reporting rule is therefore at most $0.88715560370176$.
This example illustrates the effect of the optimization error on the confidence statement even when candidate feasibility is evaluated exactly.

\section{A general certification method}
\label{sec:algorithm}
\subsection{Regionwise bounds and the computation}
A cell $B$ is a subset of the parameter space.
For each cell, suppose that we can compute
\begin{equation}
a(B)\le\inf_{\theta\in B}R(\theta),
\qquad
q_x(B)\ge\sup_{\theta\in B}G_x(\theta).
\label{eq:enclosures}
\end{equation}
We take $a(B),q_x(B)\in[0,1]$.
The values zero and one provide initial bounds for any cell.
Subsequent calculations tighten them using the structure of the model.
Cells can be rectangles, constrained subsets of rectangles or other sets for which \eqref{eq:enclosures} can be established.

Initialize $u=1$ and a finite represented covering of the full parameter space.
A candidate $\theta_c$ is accepted when a valid pointwise lower probability bound satisfies $g^-(\theta_c)>\alpha$.
If its objective is enclosed by $R(\theta_c)\le r^+(\theta_c)$, update
\[
u\leftarrow\min\{u,r^+(\theta_c)\}.
\]
Exact evaluations use $g^-(\theta_c)=G_x(\theta_c)$ and $r^+(\theta_c)=R(\theta_c)$.
The initial value $u=1$ is a universal upper bound on the target; later updates use certified feasible candidates.

The cell operations are as follows.
\begin{enumerate}
\item Exclude $B$ if $q_x(B)\le\alpha$.
\item Exclude $B$ if $a(B)\ge u$.
\item Otherwise retain $B$, evaluate optional candidates, and subdivide it when selected.
\end{enumerate}
A subdivision replaces its parent by finitely many subsets whose union covers the entire parent.
For a child $B'$ of $B$, inherit the bounds through
\[
a(B')\leftarrow\max\{a(B),a_{\rm new}(B')\},
\qquad
q_x(B')\leftarrow\min\{q_x(B),q_{\rm new}(B')\}.
\]
This retains all previously established information.
Let $\mathcal B_k$ denote the retained cells after a completed update.
Initial cells, new children and cells with updated probability bounds pass the probability-exclusion test before entering $\mathcal B_k$.
Thus every retained cell satisfies $q_x(B)>\alpha$; a cell with the initial bound $q_x(B)=1$ can await further evaluation.
The reported lower endpoint is
\begin{equation}
\ell_k=\min\bigl(\{u_k\}\cup\{a(B):B\in\mathcal B_k\}\bigr).
\label{eq:certificate}
\end{equation}
If no cells remain, $\ell_k=u_k$.

A completed update includes every child needed to replace its parent.
Stopping during a subdivision returns the last completed checkpoint.
The computation can stop when $u_k-\ell_k\le\varepsilon$ or when the available budget is used.
A best-bound strategy selects a retained cell with smallest $a(B)$.
Candidate locations, estimated probability variation and the widths of cell enclosures can help choose its subdivision.

A feasible candidate lowers the upper incumbent, while a probability enclosure can remove an entire region from the minimization.
Branch-and-bound supplies these operations \citep{hansen1980,neumaier2004}; here their validity must be checked against the ordered tail constraint and the reliability objective, since an unjustified removal could raise the reported limit above its statistical target.

\subsection{Validity at every finite budget}
\begin{theorem}[Finite-budget certificate]
\label{thm:bounds}
Suppose that \eqref{eq:enclosures} and all accepted candidate bounds are valid and that every subdivision preserves the covering.
Then, for every observation, candidate sequence and completed checkpoint,
\[
0\le\ell_k\le L_\alpha(x)\le u_k\le1.
\]
With inherited lower bounds, $\ell_k$ is nondecreasing and $u_k$ is nonincreasing.
\end{theorem}
\begin{proof}
The first exclusion rule removes a cell containing no strict-feasible parameter.
The second removes a cell whose objective is everywhere at least the current upper bound.
Subdivisions preserve coverage of the remaining parameter region.
Consequently, every strict-feasible $\theta$ with $R(\theta)<u_k$ belongs to a retained cell $B$, and
$\ell_k\le a(B)\le R(\theta)$.
For every other strict-feasible parameter, $\ell_k\le u_k\le R(\theta)$.
Taking the infimum gives $\ell_k\le L_\alpha(x)$.

The initial upper value one is valid for the $[0,1]$-valued target.
Every accepted candidate satisfies
$L_\alpha(x)\le R(\theta_c)\le r^+(\theta_c)$.
Hence incumbent updates preserve $L_\alpha(x)\le u_k$.
If the strict feasible set is empty, no candidate is accepted and the convention $L_\alpha(x)=1$ gives the same inequalities.

Children have lower bounds at least those of their parent.
Excluding a cell cannot reduce the minimum of the remaining lower bounds.
Any new upper bound is at least $L_\alpha(x)$ and therefore at least the preceding lower certificate.
These facts establish monotonicity.
\end{proof}

Theorem~\ref{thm:bounds} applies without smoothness, compactness or coordinatewise monotonicity.
Its computational requirement is the availability of valid cell and candidate bounds.
An unbounded parameter region can remain represented by cells with coarse bounds until additional analysis tightens them.
However, an unrefined retained cell with $a(B)=0$ keeps $\ell_k$ at zero; localization or a sharper exterior bound is then needed to obtain a positive limit.

\begin{corollary}[Adaptive stopping and selection]
\label{thm:coverage}
Let $S$ contain the algorithmic randomness and let $K(X,S)$ select any completed checkpoint of a valid computation.
Then
\[
\Prb_{\theta_0,S}\{\ell_{K(X,S)}(X,S)\le R(\theta_0)\}\ge1-\alpha.
\]
A finite maximum of valid certificates for the same target $L_\alpha(x)$ also has this coverage property.
\end{corollary}
\begin{proof}
Theorem~\ref{thm:bounds} gives pointwise domination for every realized $x,s$.
Thus
\[
\{\ell_{K(X,S)}(X,S)>R(\theta_0)\}
\subseteq
\{L_\alpha(X)>R(\theta_0)\}.
\]
Apply Theorem~\ref{thm:ideal}.
The maximum of certificates remains at most $L_\alpha(x)$ pointwise.
\end{proof}

Selection among computations of one prescribed ordering therefore uses their certificates directly.
If several orderings are compared after observing the data, they can be calibrated jointly.
For example, lower limits for a prespecified family of orderings $T_1,\ldots,T_M$ with noncoverage allowances $\alpha_1,\ldots,\alpha_M$ have a valid maximum whenever $\sum_m\alpha_m\le\alpha$.
This follows by the union bound.
An alternative is to include the complete ordering-selection procedure in the definition of a single statistic $T$.

\subsection{Accuracy and refinement conditions}
Coverage and convergence serve different purposes.
Theorem~\ref{thm:bounds} establishes validity at every completed step.
To establish that a narrowing interval approaches the ideal limit, we impose additional conditions on the model and refinement.

For a fixed observation, consider the boundary regularity condition
\begin{equation}
\inf_{\theta:\,G_x(\theta)>\alpha}R(\theta)
=
\min_{\theta:\,G_x(\theta)\ge\alpha}R(\theta).
\label{eq:closed}
\end{equation}
It states that equality points in the tail constraint do not reduce the objective below the strict-feasible infimum.
For continuous $R$ on compact $\Theta$, a sufficient condition is that the nonempty closed feasible set is the closure of the strict feasible set.
Under \eqref{eq:closed}, computation can equivalently use the closed constraint, accept $g^-\ge\alpha$ and exclude only when $q_x<\alpha$.

\begin{theorem}[Convergence of the certified interval]
\label{thm:convergence}
For fixed $x$, suppose that $\Theta$ is compact, $R$ and $G_x$ are continuous, the strict feasible set is nonempty and \eqref{eq:closed} holds.
Suppose that the maximum diameter of retained cells and the evaluation tolerances tend to zero, and that the objective and probability enclosures are uniformly consistent:
\[
\inf_B R-a(B)\longrightarrow0,\qquad
q_x(B)-\sup_B G_x\longrightarrow0
\]
uniformly over retained cells as their diameters and evaluation tolerances tend to zero.
Suppose also that each refined cell supplies a representative candidate whose probability lower bound and objective upper bound converge uniformly to their exact point values as the evaluation tolerances tend to zero.
Then $\ell_k\to L_\alpha(x)$ and $u_k\to L_\alpha(x)$.
Every positive width tolerance is consequently attained at some finite completed checkpoint.
\end{theorem}
\begin{proof}
If a subsequence satisfies $\ell_k\le L_\alpha(x)-\varepsilon$, its lower endpoint comes from a retained cell, since $u_k\ge L_\alpha(x)$.
Select points in those cells approaching their objective infima.
Compactness yields a convergent subsequence with limit $\theta_*$.
Shrinking diameters, continuity and enclosure consistency imply
$R(\theta_*)\le L_\alpha(x)-\varepsilon$.
A retained cell has $q_x(B)\ge\alpha$, so the same properties imply $G_x(\theta_*)\ge\alpha$.
This contradicts \eqref{eq:closed} and proves convergence of the lower endpoint.

For any $\varepsilon>0$, choose $\theta_\varepsilon$ with
$G_x(\theta_\varepsilon)>\alpha$ and
$R(\theta_\varepsilon)<L_\alpha(x)+\varepsilon$.
A cell containing this parameter cannot be probability-excluded.
If it is objective-excluded, then $u_k\le a(B)\le R(\theta_\varepsilon)$ already.
Otherwise, shrinking cells containing it supply representative candidates converging to $\theta_\varepsilon$.
The positive feasibility margin and consistent evaluations eventually certify these candidates, with objective upper bounds approaching $R(\theta_\varepsilon)$.
Hence $\limsup_k u_k\le L_\alpha(x)+\varepsilon$.
Let $\varepsilon$ decrease to zero.
\end{proof}

On compact rectangles, periodic sweeps that bisect a longest side of every surviving cell at its midpoint supply the shrinking-cover condition.
Best-bound selection can be used between these sweeps.
Subdivision rules affect efficiency, while inherited enclosures preserve validity under either selection rule \citep{csendesratz1997,neumaier2004}.
For noncompact models, analytical bounds on exterior cells or the calibrated localization in Section~\ref{sec:simulationcert} provide practical ways to organize the search.

\subsection{Simulated enclosures and parameter localization}
\label{sec:simulationcert}
A tail probability can be evaluated by simulation while retaining control over the complete adaptive computation.
For each fixed dataset $x$, let $E_x$ be the event that all bounds used for probability exclusion, objective pruning and candidate updates during the search are valid.
Suppose that
\begin{equation}
\Prb_S(E_x\mid X=x)\ge1-\delta
\quad\text{for every }x.
\label{eq:simultaneous}
\end{equation}
The event covers all completed checkpoints.
If several simulated runs are compared, the same total allowance $\delta$ covers their combined queries, through a common query index or a prespecified division of $\delta$ among runs.

A data-dependent confidence region $C(X)\subseteq\Theta$ can also localize unbounded nuisance parameters.
Assume that
\[
\inf_{\theta\in\Theta}\Prb_\theta\{\theta\in C(X)\}\ge1-\gamma.
\]
Choose nonnegative $\gamma,\delta$ with $\gamma+\delta<\alpha$, set $\alpha_0=\alpha-\gamma-\delta$, and define
\begin{equation}
L_{\alpha_0,C}(x)
=\inf_{\theta\in C(x):\,G_x(\theta)>\alpha_0}R(\theta),
\label{eq:localized}
\end{equation}
again with value one for an empty feasible set.
This uses the confidence-set principle of \citet{bergerboos1994} to account explicitly for the region excluded from the search.
The interval width in this computation measures numerical error relative to the localized target $L_{\alpha_0,C}(x)$.

\begin{theorem}[Combined statistical and computational error allocation]
\label{thm:risk}
Suppose that the cell computation over $C(x)$ is a valid lower certificate for \eqref{eq:localized} on $E_x$.
Then every adaptively selected completed output satisfies
\[
\Prb_{\theta_0,S}\{\ell(X,S)>R(\theta_0)\}
\le\alpha_0+\gamma+\delta=\alpha.
\]
The special case $C(x)=\Theta$, $\gamma=0$ gives simulation-based certification of the full-space target at level $\alpha-\delta$.
\end{theorem}
\begin{proof}
On the event
$\{\theta_0\in C(X)\}\cap\{G_X(\theta_0)>\alpha_0\}\cap E_X$,
the true parameter is feasible in \eqref{eq:localized}, and the certificate is no larger than its objective.
Noncoverage is therefore contained in the union of the three complementary events.
Their probabilities are bounded by $\gamma$, $\alpha_0$ and $\delta$, respectively.
The middle bound is the survival-transform inequality used in Theorem~\ref{thm:ideal}.
No independence among these events is needed.
\end{proof}

For an explicit construction of \eqref{eq:simultaneous} with $\delta>0$, allocate
$\delta_j=\delta/\{j(j+1)\}$ to the $j$th probability query.
Choose its parameter and positive integer batch size $m_j$ from the previous search history, then draw independent replicates conditionally on that history.
If $\widehat G_j$ is the average of its event indicators, the interval
\begin{equation}
g_j^\pm
=
\left[\widehat G_j\pm
 \sqrt{\frac{\log(2/\delta_j)}{2m_j}}\right]_{[0,1]}
\label{eq:hoeffding}
\end{equation}
covers the queried probability with conditional failure probability at most $\delta_j$ by Hoeffding's inequality \citep{hoeffding1963}.
Here $[\cdot]_{[0,1]}$ denotes clipping to $[0,1]$.
Since $\sum_j\delta_j=\delta$, both tails of all queries are simultaneously valid with probability at least $1-\delta$.
The lower endpoint can certify candidate feasibility, while the upper endpoint enters the whole-cell bounds developed below.

If sampling within a query stops adaptively, a confidence sequence replaces the fixed-batch interval.
For example, allocating $\delta_j/\{m(m+1)\}$ to sample size $m$ gives the valid radius
\[
\sqrt{\frac{\log\{2m(m+1)/\delta_j\}}{2m}}
\]
simultaneously for all positive integer $m$.
Sharper time-uniform and betting constructions can be used in the same algorithm \citep{howard2020,howard2021,waudbysmithramdas2024}.
When samples are shared across parameter queries, the enclosure must be simultaneous over the queries selected from that shared sample.
A fresh conditional batch is one simple way to meet this requirement.

\section{Constructing bounds for different reliability models}
\label{sec:enclosures}
The cell theorem uses the same two quantities, an objective lower bound and a tail-probability upper bound, across model classes.
This section gives concrete constructions.

\subsection{Monotone coherent systems}
For a coherent structure with independent component states, \eqref{eq:system} is nondecreasing in each $r_i$.
Conditioning on the remaining states gives
\begin{equation}
0\le\frac{\partial f}{\partial r_i}
=\E_{\rvec_{-i}}\{\varphi(1,Z_{-i})-\varphi(0,Z_{-i})\}\le1.
\label{eq:structurederivative}
\end{equation}
The fixed argument denotes the state of component $i$.
For independent counts $X_i\sim\operatorname{Bin}(n_i,r_i)$, use
$T(X)=f(X_1/n_1,\ldots,X_d/n_d)$.
A uniform coupling of the Bernoulli trials shows that $G_x$ is also nondecreasing in each coordinate.
Thus a rectangle $B=[a,b]$ has the exact monotone enclosures
\begin{equation}
f(a)\le f(\rvec)\le f(b),
\qquad
G_x(a)\le G_x(\rvec)\le G_x(b).
\label{eq:boxbounds}
\end{equation}

For $T(x)>0$, the finite binomial tail is continuous and $G_x(\one)=1$.
At a point with $G_x(\rvec)=\alpha$, move along $(1-s)\rvec+s\one$.
The tail is a nondecreasing polynomial in $s$ with values $\alpha$ and one at its endpoints.
It is strictly increasing: equality at two distinct arguments would make it constant on an interval and hence everywhere.
This proves the boundary regularity \eqref{eq:closed}.
For $T(x)=0$, the ideal lower limit is zero.

The closed-constraint implementation can discard a box when $G_x(b)<\alpha$.
If $G_x(a)\ge\alpha$, its lower corner is feasible and attains the minimum objective in that box; updating the incumbent to $f(a)$ then permits removal of the box.
Every other retained box has a feasible upper corner, which supplies an incumbent candidate.
These are inexpensive specializations of the general cell operations.

\begin{proposition}[Finite width for monotone boxes]
\label{prop:termination}
For a coherent reliability polynomial and the binomial ordering above, longest-side midpoint subdivision of a retained box with smallest lower objective reaches every positive interval-width tolerance in finitely many subdivisions.
\end{proposition}
\begin{proof}
For the selected box $[a,b]$, its lower corner has $f(a)=\ell$ and its feasible upper corner has already supplied an upper candidate.
By \eqref{eq:structurederivative},
\[
u-\ell\le f(b)-f(a)
\le\sum_i(b_i-a_i)
\le d\max_i(b_i-a_i).
\]
If the gap exceeds $\varepsilon$, a selected side is longer than $\varepsilon/d$.
Choose an integer $m$ such that $2^{-m}\le\varepsilon/d$.
Along a branch starting at the unit cube, longest-side midpoint subdivision halves every coordinate once in each block of $d$ subdivisions, so a node at depth $dm$ has all sides at most $2^{-m}$.
Every node selected while the gap exceeds $\varepsilon$ therefore has depth less than $dm$.
There are at most $2^{dm}-1$ such nodes in the binary subdivision tree, and a subdivision removes its selected node permanently.
Hence only finitely many subdivisions can occur before the gap is at most $\varepsilon$.
\end{proof}

As a structural example, consider a bridge network with terminals $A,D$, intermediate nodes $B,C$, and edges $AB,BD,AC,CD,BC$, numbered $1,\ldots,5$.
Conditioning on the cross edge gives
\begin{equation}
\begin{split}
f(\rvec)={}&(1-r_5)\{1-(1-r_1r_2)(1-r_3r_4)\}\\
&+r_5\{1-(1-r_1)(1-r_3)\}\{1-(1-r_2)(1-r_4)\}.
\end{split}
\label{eq:bridge}
\end{equation}
This function enters the ordering and the corner bounds directly.
The same construction covers other coherent structures through their reliability polynomials.

\subsection{Further constructions: nonmonotone structures and binomial orderings}
\label{sec:bernstein}
The remaining constructions give valid regional bounds beyond the implemented monotone-binomial case.
Their statements and proofs establish applicability, while numerical evaluation of Bernstein subdivision, continuous-model bounds, censoring and the simulation/localization branch remains separate from the experiments in Section~\ref{sec:numerical}.
For independent binomial counts, any prescribed statistic on the finite count space defines
\[
\mathcal A_x=\{k:T(k)\ge T(x)\}.
\]
Its tail probability is
\begin{equation}
G_x(\rvec)
=\sum_{k_1=0}^{n_1}\cdots\sum_{k_d=0}^{n_d}
c_k\prod_{i=1}^d
 \binom{n_i}{k_i}r_i^{k_i}(1-r_i)^{n_i-k_i},
\qquad
c_k=\one\{k\in\mathcal A_x\}.
\label{eq:exactsum}
\end{equation}
This is a tensor-product Bernstein representation.
A dense representation stores $\prod_i(n_i+1)$ tail coefficients per cell: five components with $n_i=8$ require $59049$ coefficients before subdivision.
Monotone corner evaluation avoids this per-cell coefficient array, although its tail sum and the number of retained cells can still be expensive in higher dimensions.
The basis functions are nonnegative and sum to one, so coefficient minima and maxima enclose the polynomial range.
After restriction to a subrectangle, the transformed coefficients give local probability bounds \citep{faroukirajan1988,garloff1993}.

\begin{proposition}[Bernstein cell certificate]
\label{prop:bernstein}
Let $c_k^B$ be the Bernstein coefficients of \eqref{eq:exactsum} after affine restriction to a rectangle $B$.
Then
\[
\min_k c_k^B\le G_x(\rvec)\le\max_k c_k^B,
\qquad \rvec\in B.
\]
For a binary structure $\varphi$, coherent or otherwise, the objective polynomial \eqref{eq:system} attains its minimum and maximum on $B$ at vertices.
Consequently, vertex objective bounds and Bernstein probability bounds provide the enclosures required by Theorem~\ref{thm:bounds}.
\end{proposition}
\begin{proof}
The restricted Bernstein polynomial is a convex combination of its coefficients at every point in the rectangle, proving the probability bounds.
The reliability polynomial is affine in each coordinate with the others fixed.
Its minimum and maximum can therefore be moved successively to coordinate endpoints, giving vertices of $B$.
\end{proof}

For one coordinate, de Casteljau subdivision at a fraction $\tau\in[0,1]$ uses
\[
c_j^{(0)}=c_j,\qquad
c_j^{(m)}=(1-\tau)c_j^{(m-1)}+\tau c_{j+1}^{(m-1)}.
\]
The two edges of this triangular array give the coefficient lists for the children.
Applying this operation to each tensor slice gives multivariate subdivision.
Rational coefficients and subdivision points preserve exact arithmetic.
The coefficient ranges become tight on shrinking rectangles, supplying consistent probability enclosures \citep{garloff1993}.

For example, a system that functions when exactly one of two components functions has
\[
f(r_1,r_2)=r_1+r_2-2r_1r_2.
\]
With one Bernoulli test per component, $T(X)=f(X)$ and observation $x=(1,0)$, the tail probability equals $f(\rvec)$.
Its value at $(1,1)$ is zero, while its maximum over the unit square is one.
The Bernstein and vertex bounds give the correct range $[0,1]$.
The ideal lower limit is $\alpha$, since $f$ takes every value between zero and one.
This example makes the role of regionwise bounds explicit for a nonmonotone structure.

\subsection{Continuous models through probability distances}
\label{sec:distance}
Let $\theta_B$ be a representative parameter in a cell $B$.
For total variation defined by
$\operatorname{TV}(P,Q)=\sup_A|P(A)-Q(A)|$, a bound
\[
\Delta(B)\ge\sup_{\theta\in B}
\operatorname{TV}(P_\theta,P_{\theta_B})
\]
controls every tail event generated by the chosen statistic.
If the center probability is enclosed by $[g^-(\theta_B),g^+(\theta_B)]$, then
\begin{equation}
\begin{split}
g_B^-&=\max\{0,g^-(\theta_B)-\Delta(B)\},\\
q_x(B)&=\min\{1,g^+(\theta_B)+\Delta(B)\}
\end{split}
\label{eq:tvbounds}
\end{equation}
are valid probability bounds on the whole cell.
This follows directly by applying total variation to the fixed event
$\{y:T(y)\ge T(x)\}$.

Using natural logarithms, Pinsker's inequality gives the computable choice
\begin{equation}
\Delta(B)=\min\{1,\sqrt{D(B)/2}\},
\qquad
D(B)\ge\sup_{\theta\in B}
D_{\rm KL}(P_\theta\Vert P_{\theta_B}).
\label{eq:pinsker}
\end{equation}
The divergence and variation inequalities are established tools \citep{csiszar1967,sasonverdu2016}.
Their use in \eqref{eq:tvbounds} converts a certified pointwise probability calculation into the regionwise enclosure needed for a confidence-limit certificate.
The statistic can involve parameter fitting, censoring or other prescribed data processing because the distance bound controls every measurable event.
Center probabilities can be evaluated by deterministic integration or by the simultaneous simulation construction in Section~\ref{sec:simulationcert}.

A second construction is available when the model has a simulator
$X=F_\theta(W)$ with $W$ drawn from a parameter-free law.
If a random variable $H_B^+(W)\in[0,1]$ satisfies
\[
H_B^+(W)\ge
\sup_{\theta\in B}
\one\{T(F_\theta(W))\ge T(x)\},
\]
then $\sup_{\theta\in B}G_x(\theta)\le\E H_B^+(W)$.
A concentration upper bound for this expectation is a whole-cell certificate.
Interval evaluation of the simulator and statistic can supply $H_B^+$.
This construction and \eqref{eq:tvbounds} provide two ways to control spatial variation within a cell.

\subsection{Lifetime models with retained nuisance parameters}
\label{sec:lifetime}
Consider lifetimes independent over components and repeated tests, with
\[
\log Y_{ij}\sim N(\mu_i,\sigma_i^2),
\qquad j=1,\ldots,n_i,\quad \sigma_i>0.
\]
Both location and scale remain in
$\theta=(\mu_1,\sigma_1,\ldots,\mu_d,\sigma_d)$.
At mission time $t_*$,
\[
r_i(\theta)=1-\Phi\left(\frac{\log t_*-\mu_i}{\sigma_i}\right),
\qquad
R(\theta)=f(r_1(\theta),\ldots,r_d(\theta)).
\]
A plug-in ordering estimates both parameters from each observed or replicated dataset.

For a cell center $\theta_B=(\mu_{Bi},\sigma_{Bi})_{i=1}^d$, the complete-data divergence is
\begin{equation}
D_{\rm KL}(P_\theta\Vert P_{\theta_B})
=\sum_{i=1}^d n_i
\left\{
\log\frac{\sigma_{Bi}}{\sigma_i}
+\frac{\sigma_i^2+(\mu_i-\mu_{Bi})^2}{2\sigma_{Bi}^2}
-\frac12
\right\}.
\label{eq:lognormalkl}
\end{equation}
Each summand is convex in $(\mu_i,\sigma_i)$ for $\sigma_i>0$.
On a rectangular cell, its maximum occurs at a corner.
The sum separates by component, so summing the maxima over the four endpoint pairs for each component gives $D(B)$ in \eqref{eq:pinsker}.
This supplies an explicit tail-probability enclosure with unknown location and scale.
In a certified numerical evaluation, logarithms, square roots and distribution functions in these bounds are evaluated with outward enclosures.

The component survival probability is increasing in $\mu_i$.
At fixed $\mu_i$, its direction of change in $\sigma_i$ is determined by $\log t_*-\mu_i$.
Its extrema on a parameter rectangle therefore occur at corners.
Enclosed normal-CDF evaluations give intervals $[r_i^-,r_i^+]$.
A coherent structure uses $f(\rvec^-)$ and $f(\rvec^+)$ as objective bounds; an arbitrary binary structure uses vertex bounds over the resulting reliability rectangle.

Fixed Type-I or Type-II censoring of a planned sample of $n_i$ lifetimes is a measurable function of that complete sample.
The data-processing inequality gives
\begin{equation}
D_{\rm KL}(P_\theta^{\rm obs}\Vert P_{\theta_B}^{\rm obs})
\le D_{\rm KL}(P_\theta^{\rm full}\Vert P_{\theta_B}^{\rm full}),
\label{eq:dataprocessing}
\end{equation}
so \eqref{eq:lognormalkl} also provides a valid radius for the observed-data tail \citep{csiszar1967,sasonverdu2016}.
To see this for a general measurable censoring map, write $L=dP_\theta^{\rm full}/dP_{\theta_B}^{\rm full}$.
The observed likelihood ratio is $\E_{\theta_B}(L\mid X^{\rm obs})$, and conditional Jensen's inequality for $z\log z$ gives \eqref{eq:dataprocessing}.
The statistic and replicated data use the same censoring map.
Thus the construction applies to complete and censored lognormal data with all distributional parameters retained.

For complete lognormal samples with $n_i\ge2$, exact normal-theory intervals provide one explicit compact localization.
Let $\overline Z_i$ and $S_i^2$ be the mean and unbiased variance of the log lifetimes, and allocate
$\sum_i(\gamma_{\mu i}+\gamma_{\sigma i})\le\gamma$, where every allocation is strictly between zero and one.
Use
\[
\mu_i\in
\overline Z_i\pm
t_{n_i-1,\,1-\gamma_{\mu i}/2}\frac{S_i}{\sqrt{n_i}},
\]
and
\[
\sigma_i^2\in
\left[
\frac{(n_i-1)S_i^2}{\chi^2_{n_i-1,\,1-\gamma_{\sigma i}/2}},
\frac{(n_i-1)S_i^2}{\chi^2_{n_i-1,\,\gamma_{\sigma i}/2}}
\right].
\]
Their product region has coverage at least $1-\gamma$ by the union bound and can be used in Theorem~\ref{thm:risk}.
It is compact almost surely, since $S_i>0$ almost surely under $\sigma_i>0$ and $n_i\ge2$.
A censored-data localization uses a confidence region calibrated under its observation model.

Other lifetime families enter through their full parameters in the same way.
For example, Weibull reliability is
$r_i(\eta_i,\beta_i)=\exp\{-(t_*/\eta_i)^{\beta_i}\}$, with both scale and shape in $\theta$.
A probability-distance bound, validated integration or a simulator envelope supplies its cell probabilities.
Alternatively, complete lifetime tests can be reduced to the mission-time indicators $\one\{Y_{ij}>t_*\}$.
Under independent repeated tests these give binomial counts for any lifetime distribution, and the discrete construction applies directly.
The full-lifetime and mission-indicator formulations use different amounts of information within the same certification framework.

\section{Candidate generation (optional)}
\label{sec:candidates}
Candidate search attempts to lower $u$ by finding small-reliability parameters near the tail constraint.
The finite-budget theorem permits any search routine, since the certifier checks each candidate independently before accepting it.
A score estimator, for example, uses the identity
\[
\nabla_\theta G_x(\theta)
=\E_\theta\!\left[\one\{T(X)\ge T(x)\}\nabla_\theta\log p_\theta(X)\right]
\]
when differentiation can pass through the probability integral \citep{glynn1990}.
For binomial data with $r_i=\sig(v_i)$, the logit score is $X_i-n_i r_i$, so the discontinuity of the count-space indicator does not require smoothing.

We compare primal--dual SGD, split-batch penalty SGD and sample-average optimization against the certificate alone.
Independent batches make the penalty-gradient estimate unbiased, but a stationary point of that penalty is not a global certificate.
The update equations, stationary-gradient bound, tuning values and timing table are supplied in the accompanying supplementary material.

Every strategy receives the same deterministic initialization: for a positive observed statistic, 24 bisection steps along the diagonal and each upper edge generate feasible incumbents.
The global lower calculation starts from the full parameter cube, and hybrid runs also check at most five saved stochastic candidates exactly.
This comparison asks whether the added search pays for itself by reducing the work required to reach a specified certified width.
In the four examples below it does not: final endpoints and subdivision counts are unchanged, while total time increases.

\section{Numerical studies}
\label{sec:numerical}
\subsection{Benchmark systems}
We consider four examples with $\alpha=0.1$: a scalar binomial problem and three two-component series systems.
For the latter examples, $f(r_1,r_2)=r_1r_2$.
They cover scalar inversion, a boundary optimum, an interior optimum and a partial-failure observation.

For a single component with observed count $x>0$, binomial tail inversion gives the Clopper--Pearson lower limit \citep{clopper1934},
\begin{equation}
L_\alpha(x)=\operatorname{Beta}^{-1}(\alpha;x,n-x+1),
\label{eq:scalarreference}
\end{equation}
where $\operatorname{Beta}^{-1}$ denotes the beta quantile function.
For $x=0$, the lower limit is zero.
We use $n=10$ and $x=7$.

For these examples, write $H_x(t)=\sup_{\rvec:\,f(\rvec)\le t}G_x(\rvec)$ for the largest tail probability at reliability at most $t$.
For an all-success series system, $x_i=n_i$, let $m=\min_i n_i$.
Then
\[
G_x(\rvec)=\prod_i r_i^{n_i}\le f(\rvec)^m.
\]
Equality is attained by setting all components except one with $n_i=m$ equal to one.
It follows that
\begin{equation}
H_x(t)=t^m,\qquad L_\alpha(x)=\alpha^{1/m}.
\label{eq:boundaryreference}
\end{equation}
We use $n=(6,8)$ and $x=(6,8)$.
The minimum lies on an upper edge of the parameter square.
On the diagonal, writing $t=r_1r_2$ gives $G_x=t^7$ and the larger inversion value $0.1^{1/7}\approx0.7196856730$.

For an interior example, take two equal sample sizes $n$ and $x=(1,1)$.
The tail probability is
\[
G_x(r_1,r_2)=\{1-(1-r_1)^n\}\{1-(1-r_2)^n\}.
\]
The function $h(z)=\log\{1-(1-e^z)^n\}$ is concave for $z<0$.
To verify this, put $q=1-e^z$.
Its derivative is
\[
h'(z)=\frac{nq^{n-1}}{\sum_{j=0}^{n-1}q^j}.
\]
For $n>1$, the last expression increases with $q$, since its reciprocal is proportional to $\sum_{j=0}^{n-1}q^{j-(n-1)}$.
As $q$ decreases with $z$, $h'$ is nonincreasing.
Jensen's inequality at fixed product $r_1r_2=t$ then gives
\begin{equation}
H_x(t)=\{1-(1-\sqrt t)^n\}^2,\qquad
L_\alpha(x)=\{1-(1-\sqrt\alpha)^{1/n}\}^2.
\label{eq:interiorreference}
\end{equation}
For $n>1$ and $0<\alpha<1$, the minimizing parameter is symmetric and interior.
We use $n=4$.

The partial-failure example has $n=(6,8)$ and $x=(4,7)$.
Its tail consists of count pairs satisfying $k_1k_2\ge28$.
A separate rational implementation using the same monotone box principle gives the reference interval
\[
[\,0.31508990579368396,\;0.31509000579278285\,].
\]
This reference is used as an implementation comparison at a finer tolerance of $10^{-7}$.
The first three examples provide analytical checks.

\subsection{Computational settings}
We compare the certificate alone with the three optional candidate generators, using five prespecified seeds for each generator and example.
Each SGD method has a nominal budget of $204800$ simulated count vectors; SAA--COBYLA uses a fixed bank of $2048$ simulated systems and at most $100$ objective evaluations.
The supplement specifies the update schedules, clipping rules and other tuning values.
Total time includes candidate construction, exact candidate checks, deterministic initialization and refinement, so the comparison charges each strategy for the full calculation.

The saved subdivision caps are $0$, $10$, $100$, $1000$ and $10000$, with target width $\varepsilon=10^{-5}$; refinement resumes the same state at successive caps.
The zero-subdivision checkpoint includes initialization and candidate search.
Experiments were run with Python 3.12.13, NumPy 2.4.6 and SciPy 1.17.1 on Windows 11, and the computational record retains seeds, rational endpoints, source hashes and timings.

\subsection{Accuracy and computational cost}
The record contains four problems, each run with the certificate alone and with three generators at five seeds each: 64 runs in total and 320 saved checkpoints.
All final intervals attain the width target before the 10000-subdivision cap.
Every saved interval is consistent with its analytical value or finer reference interval, and the lower endpoints are nondecreasing along each run.
For each example, all candidate generators give the same final endpoints and subdivision count.

Table~\ref{tab:accuracy} reports these endpoints.
\begin{table}[htbp]
\centering
\caption{Certified intervals at the final checkpoint, with $\alpha=0.1$ and target width $10^{-5}$.
Lower and upper endpoints are rounded outward to ten decimal places; width bounds are rounded upward from the exact rational differences.
Independent outward rounding of the endpoints can make their displayed difference exceed the width bound in the last digits.}
\label{tab:accuracy}
\small
\begin{tabular}{lrrrr}
\toprule
Case & Lower $\ell$ & Upper $u$ & Width bound & Subdivisions\\
\midrule
Scalar & 0.4482650756 & 0.4482691885 & $4.1128\times10^{-6}$ & 17\\
All-success & 0.6812827922 & 0.6812921167 & $9.3245\times10^{-6}$ & 184\\
Interior & 0.0082086771 & 0.0082186477 & $9.9705\times10^{-6}$ & 1617\\
Partial & 0.3150800582 & 0.3150900547 & $9.9964\times10^{-6}$ & 2904\\
\bottomrule
\end{tabular}
\end{table}

\begin{figure}[!htbp]
\centering
\includegraphics[width=\linewidth]{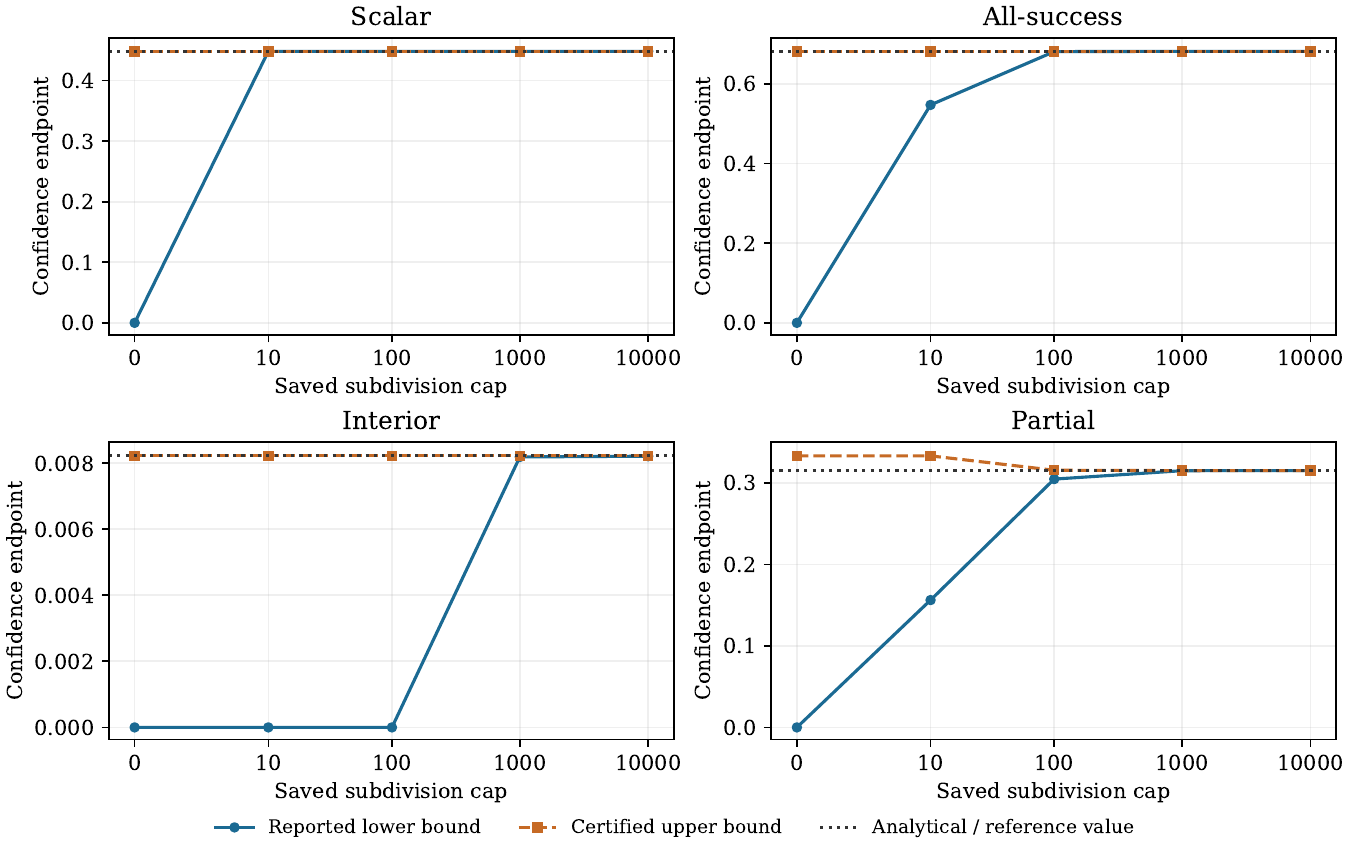}
\caption{Certified refinement for four examples using the certificate alone.
Lines join saved subdivision caps on a $\log_{10}(1+\mathrm{cap})$ horizontal scale, including initialization at zero.
Dotted lines show analytical limits for the first three cases and the midpoint of the finer reference interval for the partial-failure case.}
\label{fig:refinement}
\end{figure}

Figure~\ref{fig:refinement} shows the refinement paths for all four examples.
Increasing the subdivision budget raises the reported lower endpoint toward the ideal limit.
The distance between the endpoints gives the corresponding accuracy bound.
At a small budget, the interval is wider and the reported lower limit is more conservative.
The coverage guarantee applies throughout this refinement.

\paragraph{Cost to a given width.}
At initialization all four lower certificates equal zero, although the upper incumbents are already close to their ideal endpoints.
Reaching width $10^{-5}$ takes 17, 184, 1617 and 2904 subdivisions for Scalar, All-success, Interior and Partial, respectively.
The final recorded certificate-only times are approximately $0.89$, $5.6$, $48$ and $280$ milliseconds, including initialization and checkpoint overhead.
These are single-run descriptive timings, rather than repeated timing estimates at the first threshold crossing.
Stochastic candidate searches leave the final endpoints and subdivision counts unchanged and increase total time in every example; Supplementary Table~S1 reports the measurements.
The results favor direct refinement for these low-dimensional finite-sum problems.

\paragraph{Comparison with component limits.}
For independent components, write $c_i(\eta)$ for the component Clopper--Pearson limit in \eqref{eq:scalarreference}, with $\alpha$ replaced by $\eta$ and limit zero when $x_i=0$.
The Bonferroni product $\prod_i c_i(\alpha/d)$ is a lower confidence limit with coverage at least $1-\alpha$.
Using $\prod_i c_i(\alpha)$ instead guarantees simultaneous component coverage of at least $(1-\alpha)^d$ under independent tests, which is $0.81$ here and does not establish the requested $0.90$ level for the product limit.
Its actual product-limit coverage can be higher.
Table~\ref{tab:components} compares these limits with the returned certificate, all computed from the same observations.
The certified limit exceeds the Bonferroni product by about $63\%$ for all-success data and $119\%$ for partial failures.
This gain reflects the joint ordered construction; the certificate quantifies the small numerical loss relative to its ideal endpoint.
\begin{table}[htbp]
\centering
\caption{Component-product comparison at $\alpha=0.1$, with $n=(6,8)$.
Certificate values are rounded downward to five decimal places; component products are displayed to five decimal places.
The unadjusted column uses $90\%$ component limits and has no established $90\%$ system-level guarantee from that construction.}
\label{tab:components}
\small
\begin{tabular}{lrrr}
\toprule
Observation & Certified $\ell$ & Unadjusted product & Bonferroni product\\
\midrule
All-success $(6,8)$ & 0.68128 & 0.51090 & 0.41738\\
Partial $(4,7)$ & 0.31508 & 0.19784 & 0.14362\\
\bottomrule
\end{tabular}
\end{table}

\subsection{Complete enumeration of coverage}
For $n=(6,8)$, the sample space contains 63 observed count pairs and 28 distinct integer-product thresholds, including zero.
We compute the lower limit for every outcome using the certificate alone and PD--SGD followed by the certificate.
The latter uses three fixed seed maps, with each seed applied to every possible observation.
Each positive-product outcome receives its own candidate training and certification; a zero product returns zero directly.
Recording subdivision caps 1000 and 10000 gives $63\times4\times2=504$ outcome records.

For a true vector $\rvec_0$ and fixed seed map $s$, the coverage is
\begin{equation}
\begin{split}
C_s(\rvec_0)
={}&\sum_{i=0}^6\sum_{j=0}^8
\binom{6}{i}r_{01}^i(1-r_{01})^{6-i}
\binom{8}{j}r_{02}^j(1-r_{02})^{8-j}\\
&\hspace{3em}\times\one\{\ell_s(i,j)\le r_{01}r_{02}\}.
\end{split}
\label{eq:coverageenumeration}
\end{equation}
We evaluate both the observation weights and the event comparisons using exact rational arithmetic.
The actual returned binary64 lower limit is converted exactly to a rational before comparison.
Thus \eqref{eq:coverageenumeration} evaluates the coverage of the implemented output over the full sample space.

Table~\ref{tab:coverage} reports the results for three true reliability vectors.
The certificate alone and all three stochastic seed maps have the listed coverages at both budgets.
The 24 entries repeat the same parameter-specific results across methods and budgets: there are two distinct coverage values, $0.94078532$ and $1$, both above the nominal level $0.90$.
The coverage can stay unchanged as the interval tightens because a refined endpoint may remain on the same side of the selected true reliability for every possible observation.
The all-success observation has the largest designated statistic and hence the largest ideal lower limit, $0.681292069057\ldots$.
As this limit is below $0.72$, every outcome covers the true reliability at $(0.8,0.9)$.
These calculations check the finite-sample implementation at the stated parameter values, while Corollary~\ref{thm:coverage} applies to every true parameter.
\begin{table}[htbp]
\centering
\caption{Exact enumerated coverage at nominal confidence level $0.90$.
Each value is shared by the certificate alone and all three PD--SGD seed maps.
Displayed probabilities are rounded to eight decimal places.}
\label{tab:coverage}
\small
\begin{tabular}{lrrr}
\toprule
True components & System reliability & Cap 1000 & Cap 10000\\
\midrule
$(0.7,0.8)$ & 0.560 & 0.94078532 & 0.94078532\\
$(0.8,0.9)$ & 0.720 & 1.00000000 & 1.00000000\\
$(0.9,0.95)$ & 0.855 & 1.00000000 & 1.00000000\\
\bottomrule
\end{tabular}
\end{table}

\FloatBarrier
\section{Concluding remarks}
\label{sec:conclusion}
A Buehler limit retains its finite-sample interpretation in computation when the reported number is bounded below its ideal endpoint.
The cell certificate supplies that bound together with an upper incumbent, and its width measures the numerical loss still possible at a completed checkpoint.
Coverage survives adaptive stopping and selection among runs for the same ordering, while convergence requires additional enclosure and refinement conditions.

The general result separates the statistical model from the construction of regional bounds.
Corner evaluation gives the implemented monotone-binomial specialization; Bernstein coefficients and probability-distance inequalities extend the theory to nonmonotone and continuous models.
Unknown lognormal parameters, fixed censoring and joint allocation of localization and simulation errors are treated by explicit constructions.

Four scalar and two-component problems attain width $10^{-5}$ under all four strategies, and complete enumeration checks the implemented coverage at three parameter vectors.
The comparison with Bonferroni component limits shows the inferential gain from joint ordering, whereas the timing comparison shows that additional stochastic search has no computational benefit in these examples.
Useful deterministic incumbents leave refinement of the lower bound as the remaining work.

Dimension is the main unresolved computational constraint.
A dense binomial Bernstein representation carries $\prod_i(n_i+1)$ tail coefficients per cell, and subdivision adds a separate cost as more parameter directions must be resolved.
Path and cut representations, factorable relaxations and sharper probability-distance bounds could reduce the work per retained region.
For simulated tails, an implementation must also allocate samples across cells under the simultaneous error control of Section~\ref{sec:simulationcert}.
Evaluating these choices on bridge networks and censored lifetime data is the next computational study suggested by the theory.

\section*{Reproducibility}
The numerical results use exact rational endpoint records and complete outcome enumeration for the binomial examples.
The computational archive records the candidate seeds, feasible incumbents, subdivision and probability-evaluation counts, source hashes, package versions and elapsed times.
Tables of certified bounds use directed rounding, while the refinement figure is generated from saved checkpoints.
The supplementary material gives candidate-generation formulas, tuning values and the full timing comparison.

\begingroup
\raggedright

\endgroup
\end{document}

\maketitle
These details accompany \emph{Certified computation of Buehler confidence limits for system reliability}.
The four benchmark problems have identical final endpoints and subdivision counts across search strategies; the extra candidate searches increase total elapsed time.
Timing values below are descriptive measurements from the recorded environment, with one certificate-only run and five seeds per hybrid strategy.

\section{Stochastic candidate generation}
\label{sec:candidates}
The candidate stage explores parameter values with small system reliability near the tail constraint.
The certifier evaluates each supplied candidate and uses accepted points to improve $u$.
The finite-budget theorem applies to any candidate method, including stochastic gradients, sample-average optimization or deterministic searches.

For a regular model with density $p_\theta$, suppose differentiation can be passed through the probability integral.
With
\[
Y_x(X)=\one\{T(X)\ge T(x)\},
\qquad
s_\theta(X)=\nabla_\theta\log p_\theta(X),
\]
the likelihood-ratio identity gives
\begin{equation}
\nabla_\theta G_x(\theta)=\E_\theta\{Y_x(X)s_\theta(X)\}.
\label{eq:score}
\end{equation}
This is the standard score construction for stochastic gradient estimation \citep{glynn1990}.
When $\E_\theta s_\theta(X)=0$, an unbiased centered estimator from $b>1$ independent replicates is
\begin{equation}
\widehat d=
\frac1{b-1}\sum_{j=1}^b
(Y_j-\overline Y)s_\theta(X^{(j)}).
\label{eq:gradient}
\end{equation}
Indeed, independence across replicates gives
$\E\{\overline Y s_\theta(X^{(j)})\}=b^{-1}\E\{Y_js_\theta(X^{(j)})\}$.
A single-sample estimator uses $Y_x(X)s_\theta(X)$.

For the binomial implementation, write $r_i=\sig(v_i)$.
The logit score is $s_i=X_i-n_i r_i$, and the event is evaluated directly with its ties.
The primal--dual proposals use
\[
\mathcal L(v,\lambda)=f(\sig(v))+
\lambda\{\alpha-G_x(\sig(v))\}
\]
and
\begin{align}
v_{k+1}
&=v_k-\eta_k\{\nabla_v f-\lambda_k\widehat d_k\},\nonumber\\
\lambda_{k+1}
&=[\lambda_k+\rho_k(\alpha-\widehat G_k)]_+.
\label{eq:pd}
\end{align}
The numerical implementation clips logits and caps the multiplier.

The split-batch method uses
\[
F_\kappa(v)=f(\sig(v))+
\frac{\kappa}{2}\{G_x(\sig(v))-\alpha\}^2
\]
and two conditionally independent batches,
\begin{equation}
V_k=\nabla_v f+
\kappa(\widehat G_{A,k}-\alpha)\widehat d_{B,k},
\qquad v_{k+1}=v_k-\eta_kV_k.
\label{eq:split}
\end{equation}
Independence makes $V_k$ unbiased for $\nabla F_\kappa(v_k)$.
Section~\ref{app:sgd} gives the corresponding stationary-gradient bound under standard smoothness and variance assumptions.
These updates specialize the candidate search; their acceptance and the reported confidence limit are governed by the cell certificate.

All methods in the numerical comparison receive the same deterministic initialization.
For a positive observed statistic, 24 bisection steps on the diagonal and each upper edge produce feasible incumbents.
The global lower calculation begins with the full parameter cube.
At most five saved stochastic candidates are also checked exactly.
The comparison measures whether these additional candidates reduce the cost of reaching a specified certified width.

\subsection{Computational settings}
We compare the certificate alone with three candidate generators followed by the identical certificate: primal--dual SGD (PD--SGD), split-batch penalty SGD, and sample-average approximation with COBYLA (SAA--COBYLA).
The stochastic methods use five prespecified seeds per example.
PD--SGD uses 800 steps with batches of 256, initial primal step size $0.15$ and initial dual step size $0.3$.
Split-batch SGD uses 400 steps with two batches of 256, initial step size $0.15$ and penalty $\kappa=30$.
The step sizes are multiplied by $(1+(k-1)/500)^{-0.6}$ at step $k$.
PD--SGD clips each component probability to $[10^{-7},1-10^{-7}]$ through the logit update and caps the multiplier at 100.
SAA--COBYLA minimizes the product reliability subject to a hard-event sample-average constraint, using a fixed common-random-number bank across parameter evaluations.
The bank contains 2048 simulated systems and the optimizer has a cap of 100 objective evaluations.
The actual numbers of model evaluations are recorded for each run.

Each SGD method has a nominal budget of 204800 simulated system count vectors.
For every method, total elapsed time includes candidate construction, exact candidate checks, deterministic initialization and box refinement.
The saved subdivision caps are $0$, $10$, $100$, $1000$ and $10000$, with target width $\varepsilon=10^{-5}$.
Refinement resumes the same state at successive caps.
The zero-subdivision checkpoint includes the completed initialization and any candidate search.
The experiments were run with Python 3.12.13, NumPy 2.4.6 and SciPy 1.17.1 on Windows 11.
The recorded scripts, seeds, rational endpoints and source hashes accompany the numerical results.

\begin{table}[htbp]
\centering
\caption{Total elapsed time in milliseconds (two significant figures), including candidate generation and certification.
Hybrid entries average five seeds; the certificate alone is run once per example.}
\label{tab:time}
\small
\begin{tabular}{lrrrr}
\toprule
Method & Scalar & All-success & Interior & Partial\\
\midrule
Certificate only & 0.89 & 5.6 & 48 & $2.8\times10^2$\\
PD--SGD + certificate & 31 & 51 & $2.1\times10^2$ & $3.6\times10^2$\\
Split SGD + certificate & 26 & 46 & $2.2\times10^2$ & $3.4\times10^2$\\
SAA--COBYLA + certificate & 17 & 41 & $2.0\times10^2$ & $3.3\times10^2$\\
\bottomrule
\end{tabular}
\end{table}
\section{A stationary-gradient bound for split-batch SGD}
\label{app:sgd}
We give the standard smoothness argument for the unprojected updates \eqref{eq:split}, following the stochastic-gradient analysis of \citet{ghadimi2013}.
Suppose that $F_\kappa$ is bounded below by $F_*$, its gradient is $M$-Lipschitz, the batch size is $b$, and, conditional on the past $\mathcal F_k$,
\[
\E(V_k\mid\mathcal F_k)=\nabla F_\kappa(v_k),\qquad
\E\{\|V_k-\nabla F_\kappa(v_k)\|^2\mid\mathcal F_k\}\le\sigma^2/b.
\]
For deterministic step sizes $0<\eta_k\le1/M$, smoothness gives
\[
\E\{F_\kappa(v_{k+1})\mid\mathcal F_k\}
\le F_\kappa(v_k)
-\frac{\eta_k}{2}\|\nabla F_\kappa(v_k)\|^2
+\frac{M\eta_k^2\sigma^2}{2b}.
\]
Taking expectations, summing over $k=0,\ldots,N-1$ and using the lower bound yields
\begin{equation}
\frac{\sum_{k=0}^{N-1}\eta_k\E\|\nabla F_\kappa(v_k)\|^2}
     {\sum_{k=0}^{N-1}\eta_k}
\le
\frac{2\{F_\kappa(v_0)-F_*\}}{\sum_{k=0}^{N-1}\eta_k}
+\frac{M\sigma^2\sum_{k=0}^{N-1}\eta_k^2}
       {b\sum_{k=0}^{N-1}\eta_k}.
\label{eq:stationary}
\end{equation}
For constant step size $\eta$, the right-hand side becomes
$2\{F_\kappa(v_0)-F_*\}/(\eta N)+M\eta\sigma^2/b$.
For fixed finite sample sizes and fixed $\kappa$, the binomial probabilities are polynomials in $\rvec$, and composition with the sigmoid gives bounded derivatives.
The bounded score estimates and independent batches give variance of order $1/b$.
The step-size condition in \eqref{eq:stationary} is an assumption of this bound; the experimental step sizes are specified directly in the computational settings above.
Equation~\eqref{eq:stationary} describes stationarity of the penalty objective.
The coverage of the computed confidence limit follows from the global bounds in the finite-budget certificate theorem in the main article.